# Survivability in IP over WDM networks

Kulathumani Vinodkrishnan, Arjan Durresi, Nikhil Chandhuk, Raj Jain,
Ramesh Jagannathan, and Srinivasan Seetharaman
Department of Computer and Information Science, The Ohio State University,
2015 Neil Ave, Columbus, OH 43210-1277
Phone: 614-688-5610, Fax: 614-292-2911,
Email: {vinodkri , durresi, chandhok, jain, rjaganna, seethara }@cis.ohio-state.edu

**Abstract**

The Internet is emerging as the new universal telecommunication medium. IP over WDM has been envisioned as one of the most attractive architectures for the new Internet. Consequently survivability is a crucial concern in designing IP over WDM networks. This paper presents a survey of the survivability mechanisms for IP over WDM networks. A number of optical layer protection techniques have been discussed. They are examined from the point of view of cost, complexity and application. The case of survivability at multiple layers has been considered. The advantages and issues of multi-layer survivability have been identified. The main stressed idea is that the optical layer should provide fast protection while the higher layer should provide intelligent restoration. With this idea in mind, a relatively new scheme of carrying IP over WDM using MPLS or Multi Protocol Lambda-Switching has been discussed.

## Introduction

Challenges presented by the exponential growth of the Internet have resulted in the intense demand for broadband services. In satisfying the increasing demand for bandwidth, optical networks and more precisely WDM technologies represent a unique opportunity for their almost unlimited potential bandwidth [14, 15]. On the other hand, practically all end-user communication applications today make use of TCP/IP, and it has become clear that IP is going to be the common traffic convergence layer in communication networks. Consequently IP over WDM has been envisioned as the winning combination of the new network architecture [2].

At present, WDM is mostly deployed point-to-point and the current approaches consist in a multi-layered architecture comprising top most IP/MPLS layer over ATM over SONET/SDH over WDM, shown in Figure 1. An appropriate interface is designed to provide access to the optical network. Multiple higher layer protocols can request light paths to peers connected across the optical network. This architecture has four management layers. One can also use a packet over SONET approach, doing away with the ATM layer, by putting IP/PPP/HDLC into SONET/SDH framing. This architecture has three management layers. The most attractive proposal is the two-layer model, which aims at a tighter integration between IP and optical layers, and offers a series of important advantages over the current multi-layer architecture. The benefits include more flexibility in handling higher capacity networks, better network scalability, more efficient operations and better traffic engineering. MPLS is proposed as the integrating structure between IP and optical layers. MPLS brings two main advantages. First it can be used as a powerful instrument for traffic engineering and second it fits naturally to WDM when labels are wavelengths. An extension of the MPLS has been proposed for IP/WDM integration. This is called the Multi-protocol lambda switching which equates wavelengths to labels [3].

With the introduction of IP in telecommunications networks, there is tremendous focus on reliability and availability of the new IP-WDM hybrid infrastructures. Automated establishment and restoration of end to end paths in such networks require standardized signaling and routing mechanisms. As a result network survivability is a major criteria in comparing the solutions for integrating IP over WDM as well as the main factor in designing and operating these networks. Layering models that facilitate fault restoration are discussed. There are various proposals stating that the optical layer itself should provide restoration/protection capabilities of some form. This will require careful coordination with the mechanisms of the higher layers such as the SONET APS and the IP re-routing strategies. Hold-off timers have been proposed to inhibit higher layers backup mechanisms. Problems can also arise from the high level of multiplexing. The optical fiber links contain a large number of higher layer flows such as SONET / SDH, IP flows or ATM VCs. Since these have their own mechanisms, a flooding of alarm messages can take place. A better integration between IP and optical will provide opportunities to implement a better fault restoration [9].

Survivability refers to the ability of a network to maintain an acceptable level of service during a network or equipment failure. Mechanisms for survivability can be built at the optical transport layer or at higher network layers such as IP or ATM. The physical layer is 'close' to most of the usual faults that occur, such as a cable cut. Survivability mechanisms in the optical layer involve detecting this and performing a simple switch to divert the traffic through an alternate path. This is called protection [12]. Hence optical layer mechanisms are inherently faster. Various protection mechanisms are examined from a cost and services point of view in Section 1.

Considerations for higher layer survivability are made in Section 2. At a higher layer, an alternate path can be worked out on the basis of an algorithm, priority considerations can be made and the process can be more intelligent. This is called restoration of traffic and is more time consuming than protection [13]. Hence this mechanism cannot be used against the more common fiber cuts. It has to work along with the optical layer survivability.

Section 3 concentrates on IP over WDM integration using MPLS and the way that this architecture can handle survivability. The advantages of such architecture are examined.

## 1. Optical layer protection

Optical transport networking (OTN) is a major step in the evolution of transport networking. From an architectural point of view, the OTN is very similar to SONET/SDH. However, the major differences arise from the multiplexing technology. It is digital TDM for SDH and analog WDM for OTN. The complexities associated with analog network engineering and maintenance implications account for the majority of the challenges associated with OTN.

Survivability is central to optical networking's role as the unifying transport architecture. Survivability mechanisms at this layer are very similar those of SONET/SDH and can offer the fastest possible recovery from fiber cuts and other physical media faults. The recovery time should be in the order of 50 ms.

The optical layer can be sub divided into 2 parts, the optical channel and the optical multiplex section [1]. The optical channel corresponds to each wavelength that is being carried across the OTN. The optical multiplex section is the collection of wavelengths that arrive at an optical add/drop multiplexer in the network. Providing survivability at the channel level would be a very flexible option.

The different kinds of architectures possible are point to point, ring and arbitrary mesh and are illustrated in Figure 2.

### 1.1. Point-to-Point Mechanisms

In case of point to point, one can provide 1+1, 1:1 or 1:N protection. In 1+1, the same information propagates through 2 paths and the better one is selected at the receiver. The receiver makes a blind switch when the working path signal is poor. Unlike SONET, a continuous comparison of 2 signals is not done. 1:1 protection allows a lower priority signal to be carried along the protection fiber, as the same information is not always travelling through both fibers. A signaling channel coordinates the flow of information. 1:N protection is one in which a protection fiber is shared between N working fibers. It is usually applied for equipment protection [8].

## 1.2. Ring systems [10]

Ring mechanisms are broadly classified into: Dedicated linear protection and Shared protection rings

*Dedicated linear protection* is an extension of 1+1 protection, applied to a ring and is illustrated in Figure 3. It is effectively a path protection mechanism. Each path (Source to destination nodes) is separately protected. Since each channel constitutes a separate path, it is also called Optical channel subnetwork connection protection, (OCh-SNCP). This is usually applied to hubbed transport scenarios. For other types of connections, it is very expensive [5]. From each node, the working and protection signals are transmitted in opposite directions along the 2 fibers. At the receiving end, if the working path signal is weak, it switches to the protection path signal.

Consider the path between the OADM 2 and OADM 1. A link in the working path fails. So OADM 1 starts taking input from the protection link between B and A.

*Shared protection rings* protect a link rather than a path. Hence they are easier to setup and are the more common ring protection mechanisms.

In Figure 4 it is shown the 2 fiber SPRings case. Half the wavelengths in each fiber are reserved for protection. If these are 2 different sets of wavelengths, then there is no need for wavelength conversion, when a switch of traffic takes place between the working and protection fibers. If a link failure occurs, the OADM adjacent to the link failure bridges its outgoing channels in a direction opposite to that of the failure and selects its in coming working channels from the incoming protection channels in the direction away from the failure.

If the link between A and B fails, working traffic between A and C will now take the path ADCBC. This will be the case if only the working fiber in the link fails.

In Figure 5 it is shown the 4 fiber SPRings case. Two fibers each are allocated for working and protection. Such a system can allow span switching. Span switching means that in case of a working fiber only failing in a link, the traffic can take the protection fiber in the same span. In case of 2 fiber systems, it will have to take the longer path around the ring.

Figure 5 shows that if only the working fiber between the nodes A and B fails, there need not be a longer route.

Sometimes the need arises to protect against isolated optoelectronic failures that will affect only a single optical channel at a time. Thus we need a protection architecture that performs OCh level switching based on channel level indications. The OMS SPRing switches a group of channels within the fiber. The OCh SPRing is capable of protecting OChs independent of one another based on OCh level failure indication. An N-Channel OADM based 4 fiber ring can support upto N independent OCh SPRings.

SPRing architectures are referred to as Bidirectional line switched ring (BLSR) architectures. OCh SPRings are referred to as Bidirectional Wavelength Line Switched Ring technology, (BWLSR).
4 fiber SPRings abide by the transoceanic switching protocol defined for MS/SPRings when it performs path switching, which ensures that during a switching event, a path never traverses a span more than once, when it performs span switching. When ring switching occurs, this may not be true. This protocol is essential in long distance under sea transmissions to avoid unnecessary delay.

**1.3. Mesh Architectures**

Along a single fiber, any two connections cannot use the same wavelength. The whole problem of routing in a WDM network with proper allocation of a minimum number of wavelengths is called the routing and channel assignment problem. It is found that in arbitrary mesh architctures where the connectivity of each node is more, the number of wavelengths required greatly decreases. This is the advantage of having a mesh architecture. Moreover addition of new nodes and removing existing nodes becomes very easy.

Finding an alternate path everytime a failure occurs would be a time consuming process. Hence an automatic protection switching mechanism, like that for the rings, is required. Three alternatives are briefly discussed here:

**Ring Covers**

The whole mesh is divided into smaller cycles in such a way that each edge comes under atleast one cycle. Along each cycle, a protection fiber is laid. It may so happen that certain edges come under more than one cycle. In these edges, more than one protection fiber will have to be laid. Hence, the idea is to divide the graph into cycles in such a way that this redundancy is minimized [16]. However, in most cases the redundancy required is more than 100%.

**Protection Cycles [4]**

This method reduces the redundancy to exactly 100%. The networks considered have a pair of bi-directional working and protection fibers. Fault protection against link failures is possible in all networks that are modeled by 2 edge connected digraphs. The idea is to find a family of directed cycles so that all protection fibers are used exactly once and in any directed cycle, a pair of protection fibers is not used in both directions unless they belong to a bridge.

For planar graphs, such directed cycles are along the faces of the graph. For non planar graphs, the directed cycles are taken along the orientable cycle double covers which are conjectured to exist for every digraph. Heuristic algorithms exist for obtaining cyclic double covers for every non planar graph.

The maximum number of failures that such an architecture can protect with exactly 100% redundancy, given that no 2 failures occur in the same protection cycle, is F/2 rounded to lower integer, where F is the number of faces for the planar graph. For a non planar graph, the number is (n-1)/2 where 'n' is the order of the graph.

Figure 6 shows five protection cycles formed along the faces of the planar graph. The 2 directions of link 'a' are protected by cycles 1 and 2. The 2 directions of link 'b' are protected by cycles 3 and 5. Thus, link protection is brought about.

**WDM Loop back recovery [11]**

A double cycle ring cover covers a graph such that each edge is covered by two cycles. Cycles can then be used as rings to perform restoration. For a two edge connected planar graphs, a polynomial time algorithm exists which can create double cycle covers. For two edge connected non planar graph, the existence of a double cycle cover is a conjecture and no algorithm other than an exhaustive search exists.

We cannot assign primary and secondary wavelengths in such a way that a wavelength is secondary or primary over a whole ring. Therefore the cycle double cover would require wavelength changing which is not feasible.

**General method of selecting directions and Performing WDM Loop-back for link failures**

An undirected graph G = (N, E) is a set of nodes N and edges E. With each edge [x,y] of an undirected graph, we associate arcs (x,y) and (y,x) . A directed graph P = (N, A) is a set of nodes N and directed arcs A. Given a set of directed arc , A, define the reversal of A to be A' = { (i,j) | (j,i) ε A }. Similarly, given any directed graph P = (N, A) define P' = (N, A') to be the reversal of P.

Given a two-vertex (edge)-connected graph or redundant graph G = (N,E ) i.e removal of a vertex leaves the graph connected. The idea is to create a pair of spanning sub-graphs, B = (N,A) and R = B' = (N,A') each of which can be used for primary connections between any pair of nodes in the graph. In the event of a failure, connections one wavelength in B are looped back on the same wavelength around the failure using R.

Consider the example shown in Figure 7. If [y,z] were to go down , then all the traffic going from y to z would use the graph R and go from y ->x -> w -> z . Then z will use graph B itself to route the traffic as it would do if the traffic would have arrived from x. Now for traffic coming fro z to y, B is used and traffic is routed z-> w -> x -> y. Edge [y,z] can be successfully bypassed if there exists a path with sufficient capacity from y to z in R and a path with sufficient capacity exists from z to y in B. Details of determing B, given a graph G can be found in [6].

Advantages of WDM Loop back recovery includes:

1. Allows two-fiber restoration.
2. Has a simple polynomial time algorithm regardless of the planarity of the graph, which describes the network topology.
3. Allows different choices of routing for backup connections in case of restoration.

A Summary of point-to-point and ring architectures is shown in Table 1.

## 2. Consideration for Multi-layer survivability

The optical layer shall form the lower most layer of the network architecture. It provides the service of transport to higher layers. The issues are choice of the higher layer and the presence of survivability mechanisms in one or more of the higher layers.

Following are the problems if only the optical layer provides survivability:

- It cannot handle higher layer equipment failures
- It cannot monitor high bit error rates and cause protection switching
- The granularity of protection is not fine enough. It cannot provide different levels of protection to different parts of the traffic.
- It does not provide an optimized alternate path. The protection path may be very long.

However, there exist complications with multi-layer survivability.

- A failure at the optical layer can trigger off multiple alarms at the higher layer. For e.g., a single fiber may contain a number of SONET streams. Its failure can cause an alarm explosion.
- All the layers try to rectify the same fault, causing chaos.
- There is potential for considerable wasted bandwidth resulting from each layer having its own spare resources to use during faults.

Multi-layer recovery can combine the merits of optical layer and the higher layer schemes. More specifically, the protection mechanism of optical layer can be combined with the restoration mechanism of the higher layers. A good strategy would be the nesting of survivability mechanisms and avoidance of unnecessary interworking. In such a case, combining SONET protection with optical layer protection would be a waste. Instead an intelligent higher layer recovery can be incorporated.

To achieve the objective of a closer IP-WDM integration, an MPLS-based approach named as Multi protocol Lambda switching has been suggested. The wavelengths have been equated with labels. This restoration mechanism and how it can operate with optical layer protection, is discussed in the next section.

## 3. Survivability in IP over WDM using MPLS architecture [3]

The model consists of IP routers attached to an optical core network. The optical network consists of multiple Optical Cross-Connects (OXCs), interconnected by optical links. Each OXC is capable of switching a data stream using a switching function, controlled by appropriately configuring a cross-connect table [9]. The switching within the OXC can be accomplished either in the electrical domain, or in the optical domain. In this network model, a switched optical path is established between IP routers.

Multiprotocol Label switching is a switching method in which each hop by hop decision is based on a label. The ordered set of labels is called a Label switched Path (LSP). Each LSP has a set of criteria associated with it, which describes the traffic that traverses the LSP. A set of criteria groups the traffic into Forwarding Equivalence Class (FEC). FECs group IP traffic together and associate with them what is commonly called a next hop: a tuple that defines the interface and the low level IP address of the next IP router. LSPs are setup using signaling protocols like [RSVP] or [CR-LDP]. A device that can group traffic into an FEC is called a Label edge router while as a device, which bases its decision. An LSR performs label switching by establishing a relation between an <input port, input label> tuple and an <output port, output label> tuple. Similarly, OXC provisions optical channel trail by establishing a relation between an <input port, input optical channel> tuple and an <output port, output optical channel> tuple. Consider an arbitrary mesh connection of optical routers (OXCs) shown on Figure 8.

Protection switching relies on rapid notification of failures. Once a failure is detected, the node that detected the failure must send out a notification of the failure by transmitting a Failure Indication signal (FIS) to those of its upstream LSRs that were sending traffic on the working path that is affected by the failure. This notification is relayed hop-by-hop by each subsequent LSR to its upstream neighbor, until it eventually reaches a Path Switched LSR, (PSL)

The PSL is the LSR that originates both the working and protection paths, and the LSR that is the termination point of both the FIS and the FRS. Note that the PSL need not be the origin of the working LSP.

The Path Merge LSR (PML) is the LSR that terminates both the working path and its corresponding protection path. Depending on whether or not the PML is a destination, it may either pass the traffic on to the higher layers or may merge the incoming traffic on to a single outgoing LSR. Thus, the PML need not be the destination of the working LSP.
An LSR that is neither a PSL nor a PML is called an intermediate LSR. The intermediate LSR could be either on the working or the protection path, and could be a merging LSR (without being a PML).

Since the LSPs are unidirectional entities and protection requires the notification of failures, the failure indication and the failure recovery notification both need to travel along a reverse path of the working path from the point of failure back to the PSL(s). When label merging occurs, the working paths converge to form a multipoint-to-point tree, with the PSLs as the leaves and the PML as the root. The reverse notification tree is a point-multipoint tree rooted at the PML along which the FIS and the FRS travel, and which is an exact mirror image of the converged working paths

The establishment of the protection path requires identification of the Working path, and hence the protection domain. In most cases, the working path and its corresponding protection path would be specified via administrative configuration, and would be established between the two nodes at the boundaries of the protection domain (the PSL and PML) via explicit (or source) routing using LDP , RSVP, signaling (alternatively, using manual configuration).

The Reverse Notification Tree (RNT) is used for propagating the FIS and the FRS, and can be created very easily by a simple extension to the LSP setup process, see Figure 9. During the establishment of the working path, the signaling message carries with it the identity (address) of the upstream node that sent it. Each LSR along the path simply remembers the identity of its immediately prior upstream neighbor on each incoming link. The node then creates an inverse cross-connect table that for each protected outgoing LSP maintains a list of the incoming LSPs that merge into that outgoing LSP, together with the identity of the upstream node that each incoming LSP comes from. Upon receiving an FIS, an LSR does the following:

- Extracts the labels contained in it (which are the labels of the protected LSPs that use the outgoing link that the FIS was received on)
- Consults its inverse cross-connect table to determine the identity of the upstream nodes that the protected LSPs come from
- Creates and transmits an FIS to each of them.

The reverse notification tree arising from node 8 is shown in the figure. When the link between 5 and 8 fails, the FIS starts from 5. At node 4, it will branch to 6 and 3. [7].

The advantages of the above architecture include:

- This is much faster than layer 3 IP restoration

- They can easily provide path protection with different priorities for each path. A finer granularity of control is established.
- Features like Quality of Service can be easily brought about
- The above operations will be timer induced. So there will be no problems in the inter operation with the optical layer.
- MPLS can incorporate a fiber level protection scheme to improve the scalability of WDM survivability schemes by using the MPLS stack function. All LSP labels are merged into a larger, fiber LSP. A full fiber LSp is switched to a similar backup LSP, which mirrors the individual LSP label assignments. Restoring the encapsulated fiber LSP improves response times and decreases fault related MPLS signaling.

## Conclusion

Various survivability techniques at the optical layer of IP over WDM architecture were presented. These were essentially Automatic protection Switching mechanisms. They are fast and of the order of 50 ms. The need to add intelligence into the process and establish a fine grain of control by setting up priorities and bringing quality of service into survivability is felt. For this, multi layer survivability is suggested. Reliance on SONET in WDM systems inhibits growth of packet over fiber technology. Restoration mechanisms at the IP layer are extremely slow. Hence IP over WDM using lambda switching is considered. This offers restoration through fast signaling. Priorities can be added. Interworking with the WDM layer can be regulated using timers. Thus, protection at the optical layer and restoration at the MPLS layer is suggested.

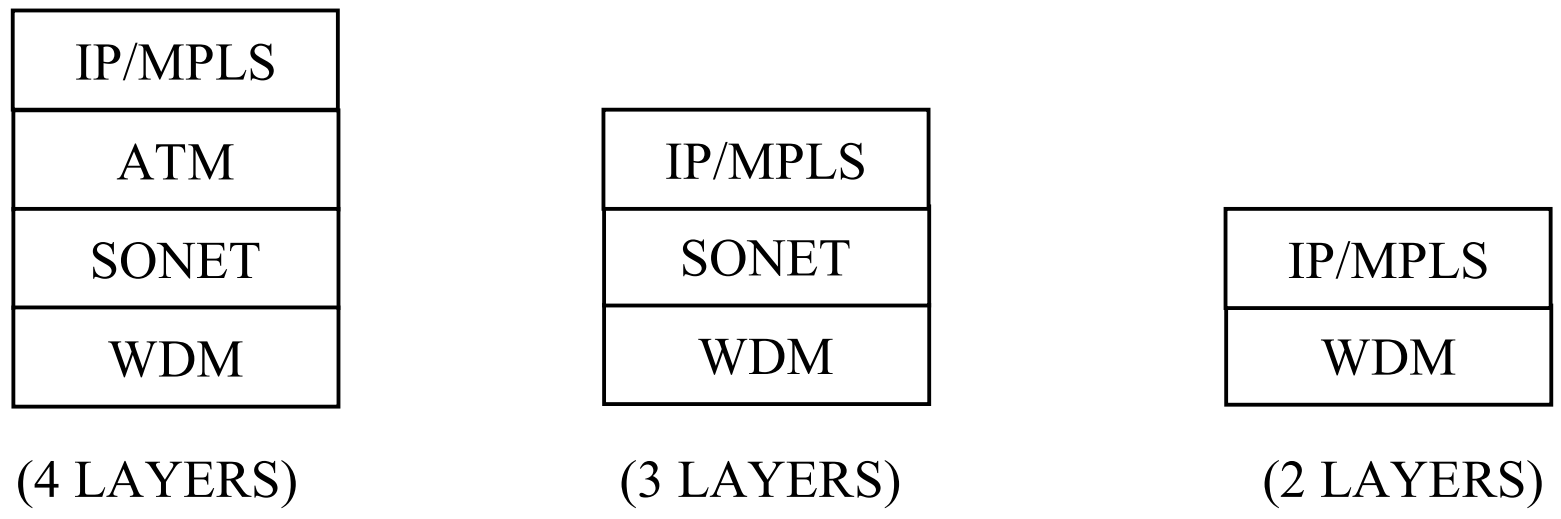

Figure 1. Layering Architectures Possible

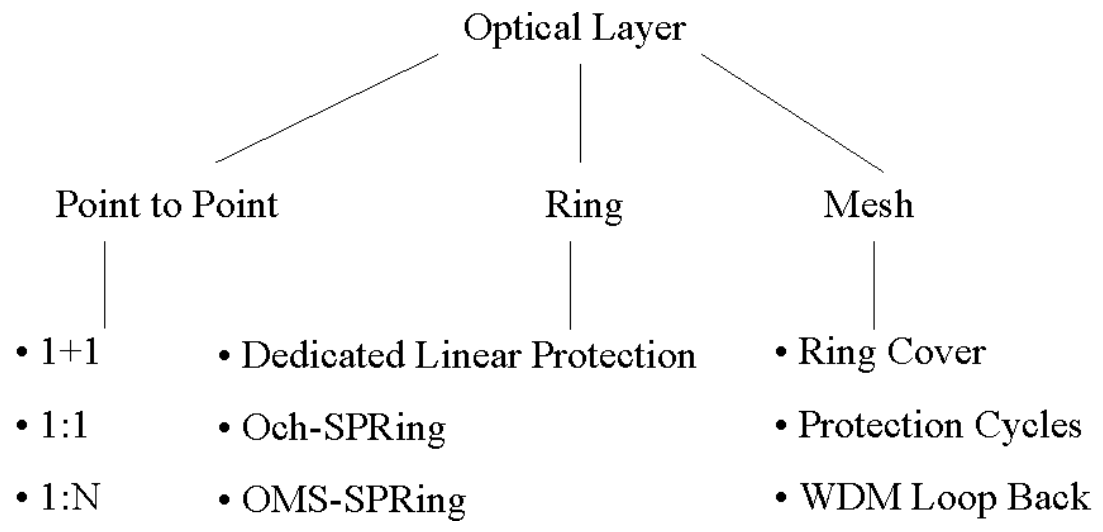

Figure 2. . Optical Layer Protection Mechanism

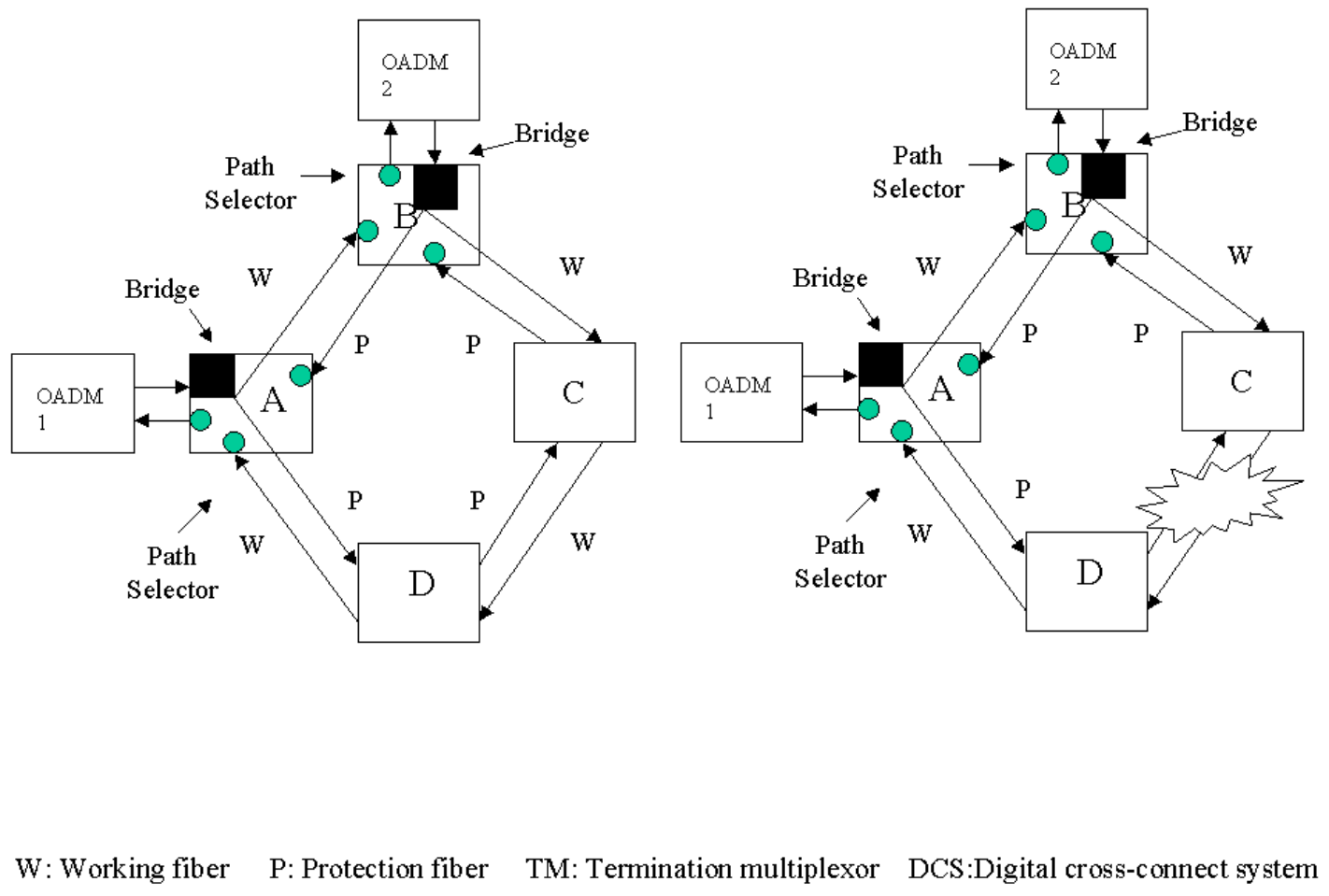

Figure 3. Dedicated Linear Protection

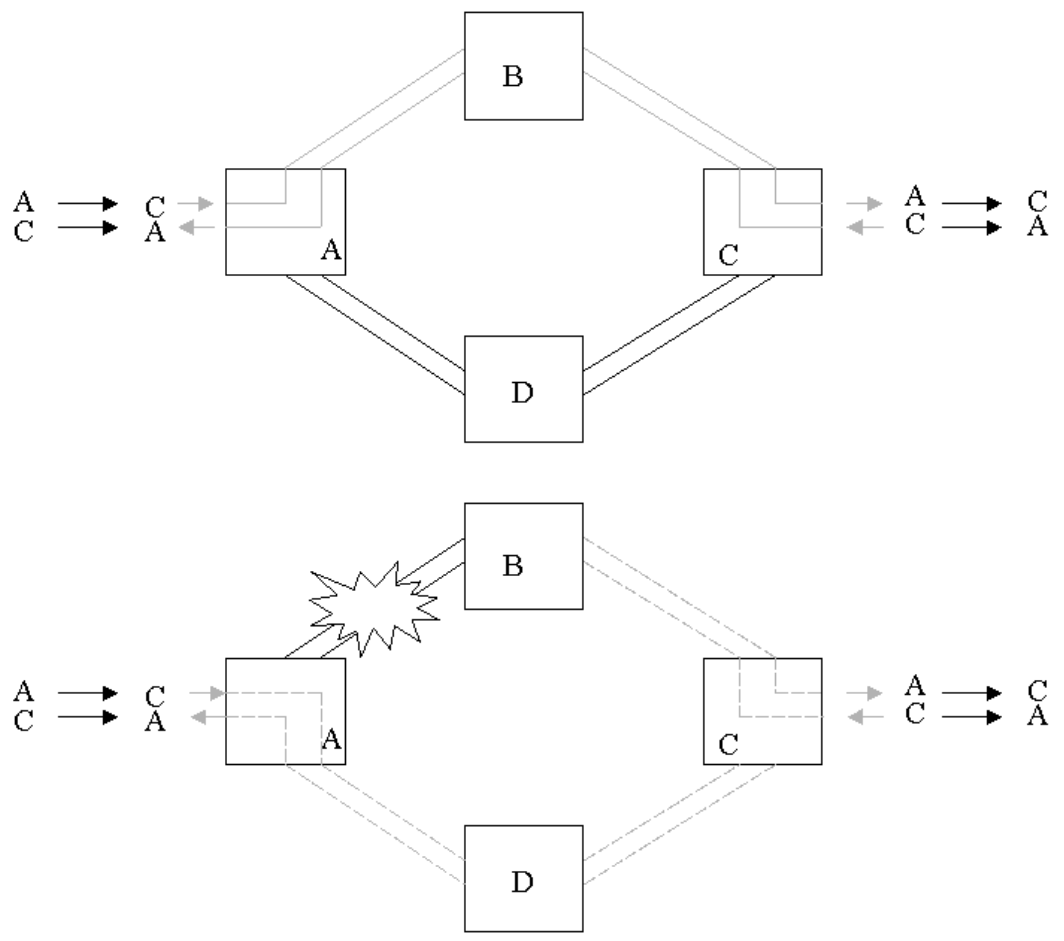

Figure 4. 2 Fiber SPRing - Showing Ring Switching

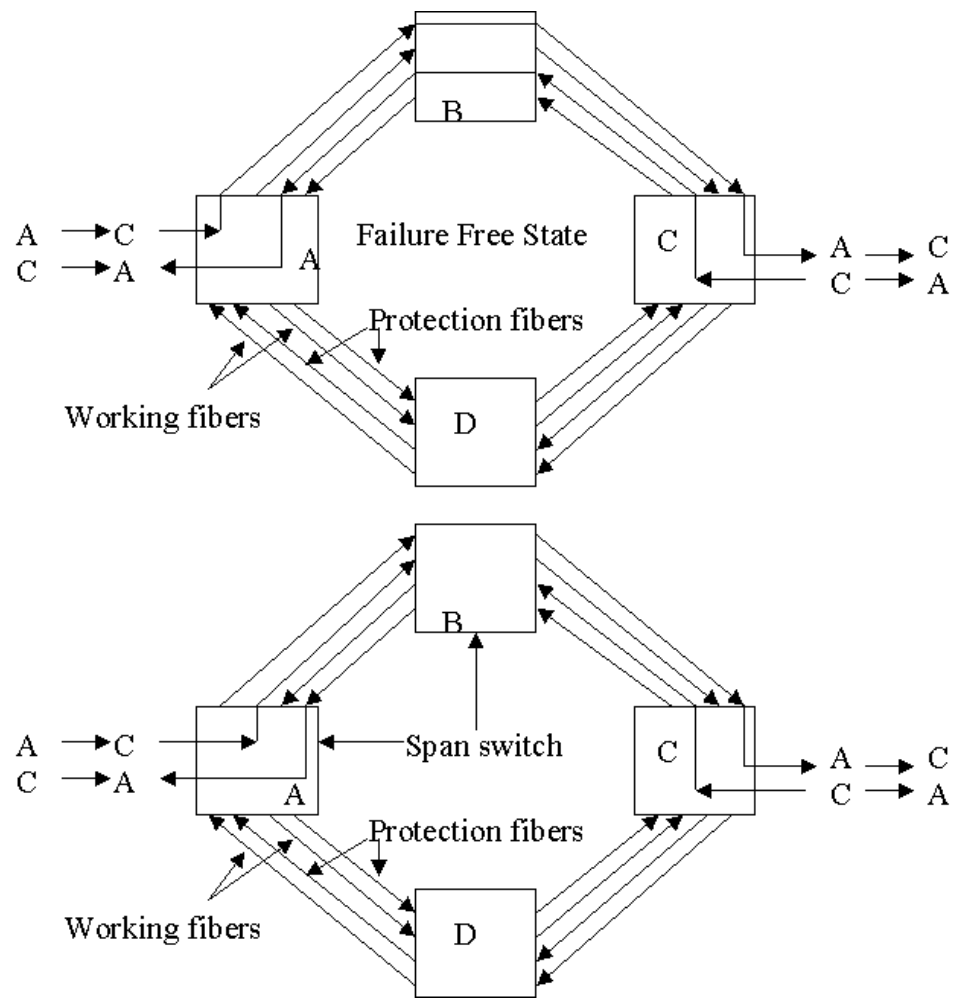

Figure 5. 4 fiber SPRing - Showing Span Switching

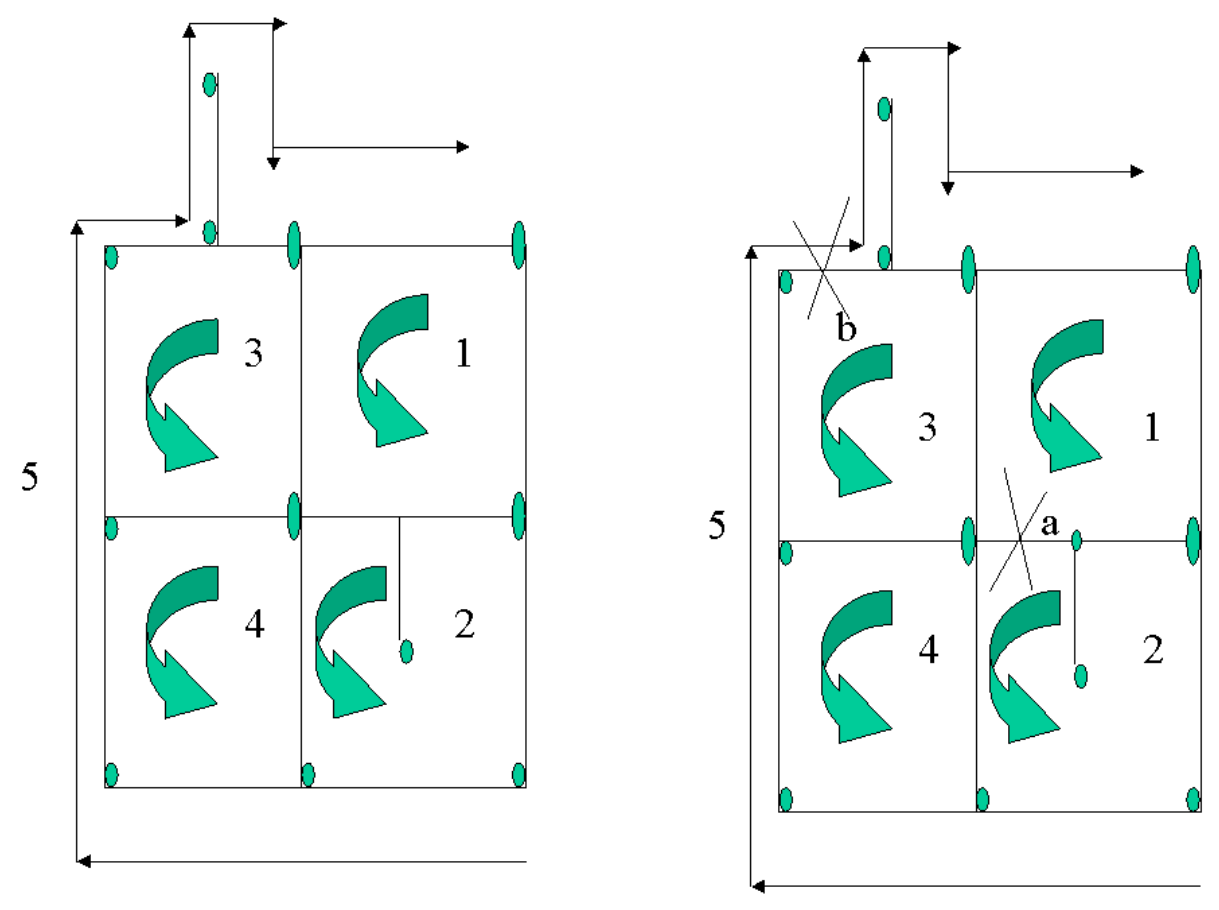

Figure 6. Protection Cycles in a planar graph

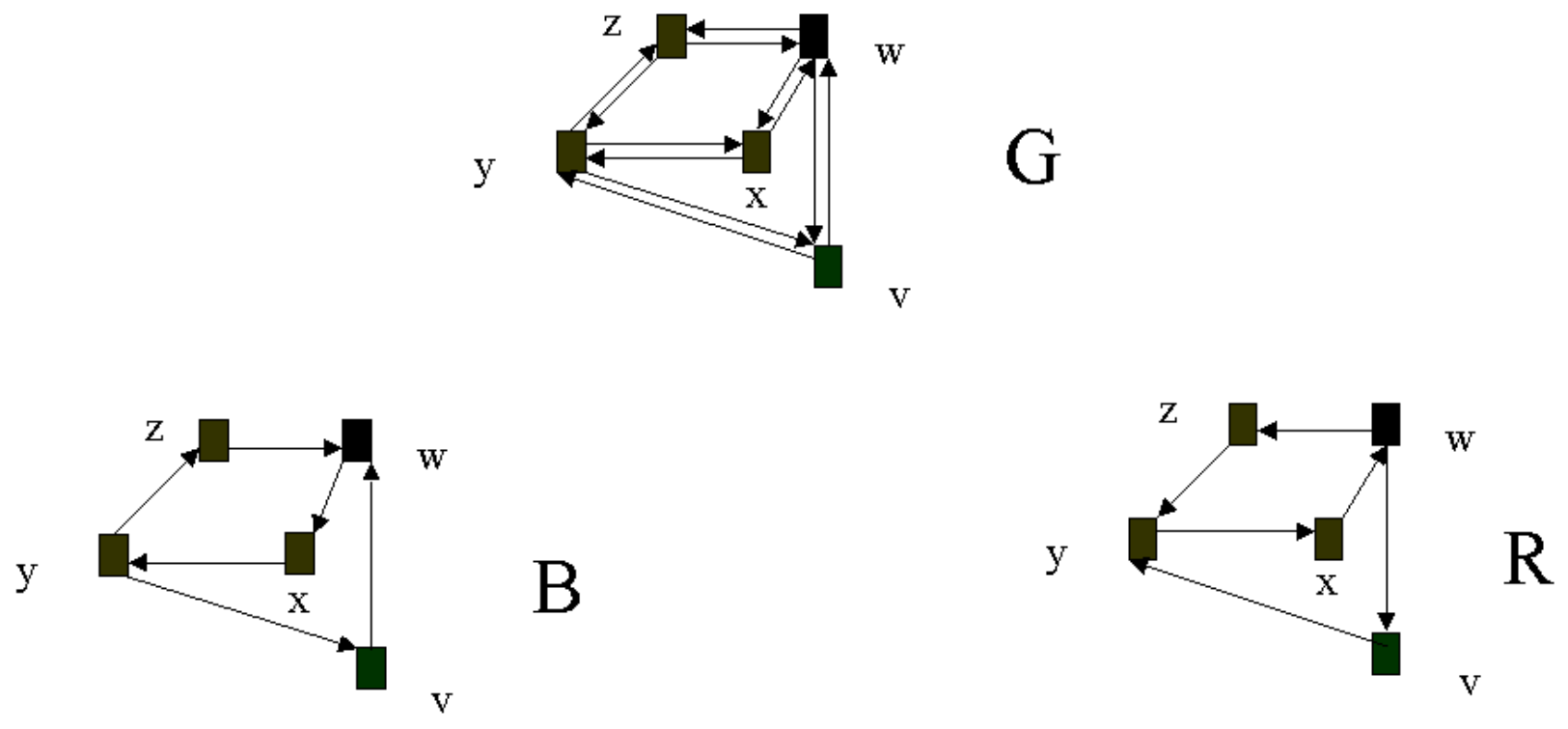

Figure 7. Illustration of WDM loop back using graph G (B and R are subgraphs of G)

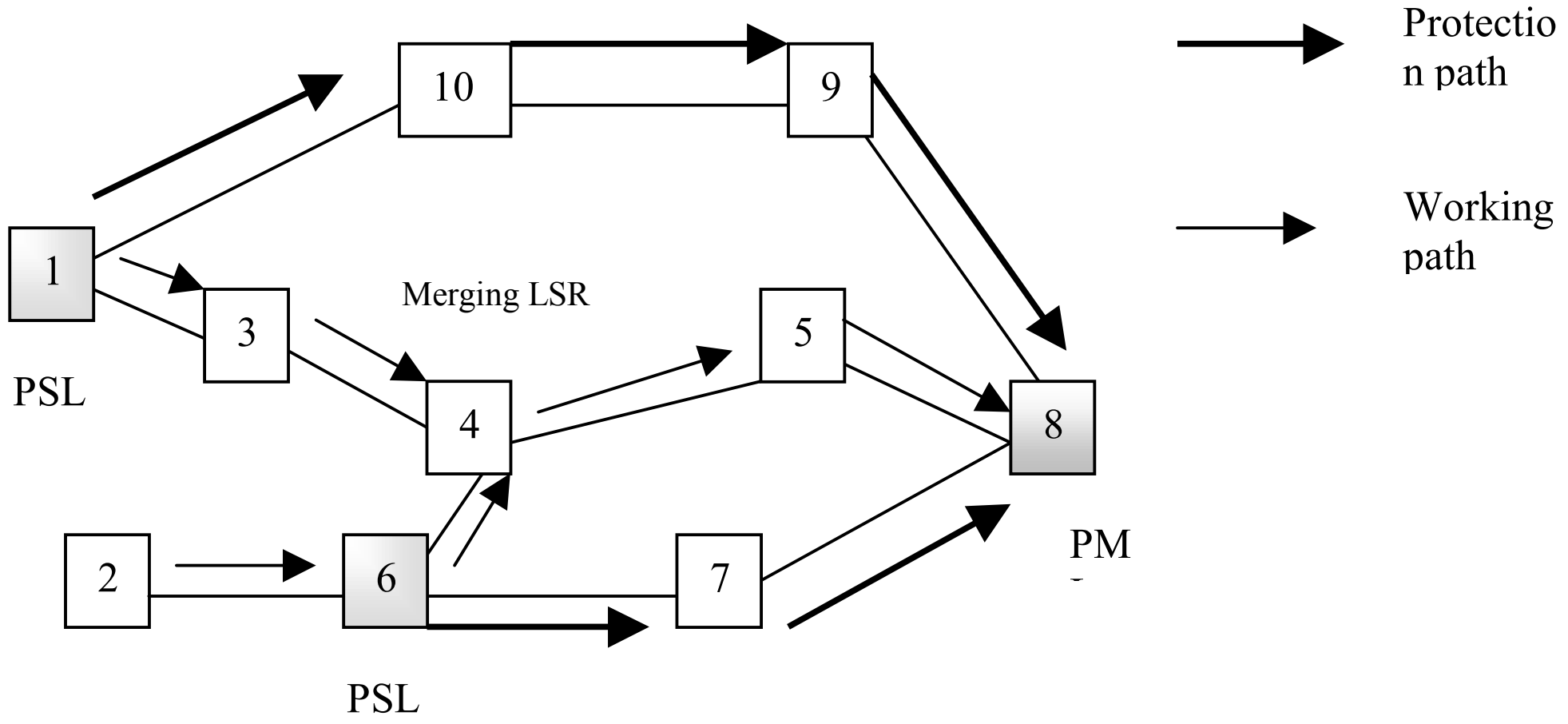

Figure 8. A mesh connection of optical routers

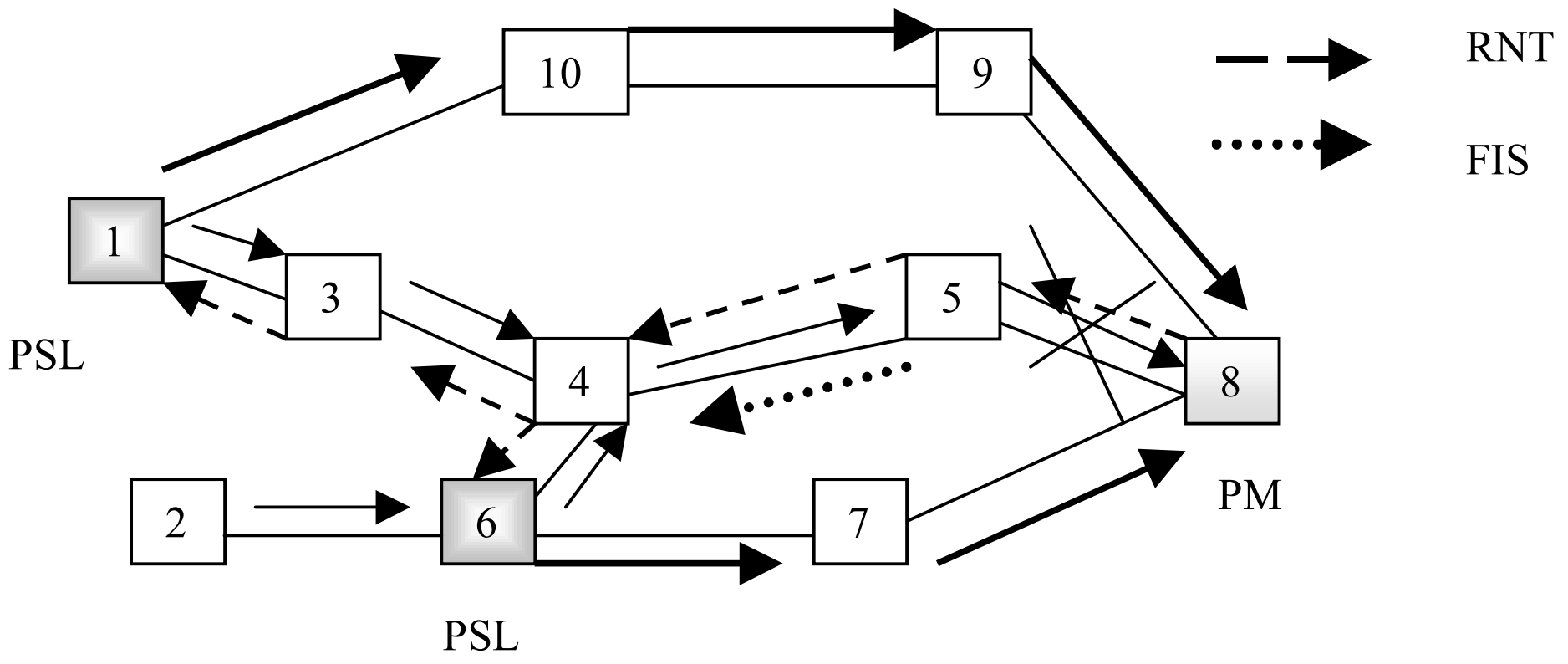

Figure 9. FIS travelling along Reverse Notification Tree

Table 1. Comparing point-to-point and ring architectures

| | |
|---|---|
| 1+1 Path Protection | When applied to a ring, called as UPSR<br>Bandwidth inefficient<br>Requires diverse paths in the network<br>Easy implementation<br>Very costly, as separate bridges required for each path<br>Useful in hubbed network implementations |
| 1:1 Line protection | When applied to a ring, called as BLSR<br>Can be done at a channel level (Called as BWLSR)<br>More Bandwidth Efficient<br>Low priority traffic supported<br>More useful in rings |